\documentclass[12pt,preprint]{aastex}

\begin{document}

\title {THE MASSIVE DISK AROUND OH 231.8+4.2}

\author{M. Jura, C. Chen and P. Plavchan} 
\affil{Department of Physics and Astronomy, University of California,
    Los Angeles CA 90095-1562; jura@clotho.astro.ucla.edu; cchen@astro.ucla.edu; plavchan@astro.ucla.edu}

\begin{abstract}
We have obtained 11.7 ${\mu}$m and 17.9 ${\mu}$m images at the Keck I telescope of the circumstellar dust emission from OH 231.8+4.2, an evolved mass-losing
red giant with a well studied bipolar outflow.  We detect both a central unresolved point source with a diameter less than 0{\farcs}5 producing F$_{\nu}$(17.9 ${\mu}$m) = 60 Jy, and  emission extended more than 1{\arcsec} away from the star
which is aligned with the bipolar outflow seen on larger scales.   We find that  the unresolved central source can be explained by an opaque,  flared disk with an outer radius of ${\sim}$5 ${\times}$ 10$^{15}$ cm and an outer temperature of ${\sim}$130 K.    One possible model to explain this flaring is that the material in the disk is orbiting the central star and not simply undergoing a radial expansion.       
\end{abstract}

\section{INTRODUCTION}
Solar-type stars  typically lose over half their
initial main sequence mass on the Asymptotic Giant Branch (AGB) before becoming white dwarfs (Habing 1996); mass loss  is 
of central importance both in stellar evolution and the replenishment of
interstellar matter. 
While many  AGB stars exhibit approximately spherically symmetric mass loss which can be understood in terms of standard models of the interaction of stellar pulsations and radiation pressure on dust (Lamers \& Cassinelli 1999, Willson 2000),
others show markedly asymmetric outflows which are not  well understood.  
Here, we report an investigation  to learn more about the circumstellar material around OH 231.8+4.2, a star with a distinctive bipolar outflow.

OH 231.8+4.2 appears to be an evolved member of the open cluster M 46 (Jura \& Morris 1985), and on the basis of the main sequence turnoff from this system, this mass-losing star probably had an initial
main sequence mass of about 3 M$_{\odot}$.  Both
its cluster membership  and  ``phase-lag" measurements in
the reflection nebulosity  (Kastner et al. 1992) indicate the distance to the star
is approximately 1.3 kpc. Although the star is  buried in an optically opaque circumstellar dust cloud, 
scattering in bipolar lobes perpendicular to the nominal disk  indicates the presence of a luminous M9 star (Cohen \& Frogel 1977).   The star is variable with a  period which might be as short as 648 days (Feast et al. 1983) or as long as 708 days (Kastner et al. 1992), and its luminosity ranges  between 10$^{4}$ L$_{\odot}$ and
2 ${\times}$ 10$^{4}$ L$_{\odot}$ (Kastner et al. 1998). The luminosity and effective temperature can be explained if the star lies on the AGB.

At 60 ${\mu}$m, OH 231.8+4.2 is brighter than every other AGB star in
the sky except IRC +10216 and the Egg Nebula (RAFGL 2688) (see Jura \& Kleinmann
1989, Kleinmann et al. 1978). IRC+10216 is very bright at 60 ${\mu}$m because its distance is only ${\sim}$150 pc; it is not uniquely luminous.  For OH 231.8+4.2, however, with an estimated distance of 1.3 kpc, then as found in the eDIRBE data from the COBE satellite, $L_{\nu}$(60 ${\mu}$m) can be as large as 3 ${\times}$ 10$^{24}$ erg s$^{-1}$ Hz$^{-1}$ which
is comparable to that from the Egg Nebula and, as shown in Figure 1,  substantially  larger
than $L_{\nu}$(60 ${\mu}$m) displayed by most other mass-losing AGB stars.

The optical and near-infrared morphology of OH 231.8+4.2 indicates that there is a massive disk surrounding the
central star (see, for example, Kastner al. 1992).    
Additionally,  OH 231.8+4.2  displays a 
bipolar outflow at radio and optical wavelengths which is inferred to carry at least 0.3 M$_{\odot}$ (Reipurth 1987, Morris et al. 1987, Sanchez-Contreras et al. 2000, Alcolea et al. 2001).  The bipolar morphology exhibits discrete clumps  (Alcolea et al. 2001).   

Although not directly determined spectroscopically, in order to account for the bipolar morphology, a plausible assumption is that
OH 231.8+4.2 is a  binary and thus the outflow may be very flattened (Mastrodemos \& Morris 1999, Soker \& Rappaport 2000, Soker 2002).  The nature of the putative companion is unknown.   It is unlikely that the companion is a white dwarf since  OH 231.8+4.2 does not exhibit any spectroscopic similarity to 
 symbiotic systems.
 
The dynamical nature of the disk is not known.  Is the matter expanding
radially outwards from the star or is it gravitationally bound and
orbiting the star, as appears to be the case for some other evolved red giants
with circumstellar envelopes such as the Red Rectangle (Jura \& Kahane 1999)?  
In order to learn more about OH 231.8+4.2, we have obtained high angular resolution mid-IR images
of this system with the Keck I 10m telescope.  Previously,  Meixner et al. (1999) have obtained mid-IR images of this system, but with a 4m telescope.
With the larger aperture of the Keck telescope, we can study more compact structures.  Many other
studies have focused on the large scale bipolar outflow; here we
concentrate on the inner region of the system.

\section{OBSERVATIONS}

Our data were obtained on  2001 Feb 05 (UT)  at the Keck I telescope 
using the Long Wavelength Spectrometer (LWS) which was built by a team 
led by B. Jones and is described on the Keck web page.  The LWS is a 
128 ${\times}$ 128 SiAs BIB array with
a pixel scale at the Keck telescope of 0{\farcs}08 and a total field of 
view of 10{\farcs}2 ${\times}$ 10{\farcs}2.  We used the ``chop-nod" 
mode of observing, and 2 different filters centered at 11.7 ${\mu}$m and 17.9 ${\mu}$m with widths of 1.0 ${\mu}$m and 2.0 ${\mu}$m, respectively.  Following Chen \& Jura (2001), we used Capella (= HR 1708)  for flux and point-spread-function calibrations.  For Capella the FWHM of the image was 0{\farcs}47 and 0{\farcs}49 at 11.7 ${\mu}$m and 17.9 ${\mu}$m,  respectively.   

The images in the 11.7 ${\mu}$m in the 17.9 ${\mu}$m  filters are presented in 
Figures 2 and 3. The Keck data show  an extension 
at position angle ${\sim}$22$^{\circ}$ that has a full length of ${\geq}$3{\arcsec}.  The extended infrared emission which was reported by Meixner et al. (1999)  shows the same orientation in the sky
as the bipolar lobes detected at other wavelengths.  There is also a central unresolved point-like source.
We measure total fluxes of F$_{\nu}$(11.7 ${\mu}$m) = 27 Jy and
F$_{\nu}$(17.9 ${\mu}$m) = 240 Jy.  These fluxes are about 50\% greater than the average values
determined by IRAS of F$_{\nu}$(12 ${\mu}$m) = 19 Jy and F$_{\nu}$(25 ${\mu}$m) = 226 Jy.  However, OH 231.8+4.2 is strongly variable at mid-infrared wavelengths. The DIRBE instrument on the COBE satellite measured
 mean fluxes of F$_{\nu}$(12 ${\mu}$m) = 72 Jy and F$_{\nu}$(25 ${\mu}$m)
= 680 Jy.   Thus our measured fluxes  are within the range
given by previous observations. If the pulsational period given by Kastner et al. (1992) has remained stable, then our data were obtained near
minimum light.   Both the COBE and IRAS beams are much larger than the
angular resolution that we obtained, but the fluxes reported with these instruments may still be comparable
to our results since we detect emission as far as 2${\arcsec}$ from the
central source.  At this angular offset,  we expect the grain temperature to have fallen to
170 K (see below) and thus there would probably be relatively little
extended emission beyond this region at 11.7 ${\mu}$m and even 17.9 ${\mu}$m.
We find that the unresolved central
sources carry about 1/3 and 1/4 of the total flux at 11.7 ${\mu}$m and 17.9 ${\mu}$m, respectively, so that within the unresolved core, F$_{\nu}$(11.7 ${\mu}$m) = 9 Jy and
F$_{\nu}$(17.9 ${\mu}$m) = 60 Jy.

It is notable that OH 231.8+4.2 does display a bright unresolved central source.
Other stars with ratios of F$_{\nu}$(25 ${\mu}$m)/F$_{\nu}$(12 ${\mu}$m) ${\geq}$ 10 in the IRAS data, indicative of cold grains, show distinct shells  and no central  dust source. For example, HD 179821, with F$_{\nu}$(25 ${\mu}$m)/F$_{\nu}$(12 ${\mu}$m) = 21 shows a resolved shell with an inner
diameter of about 3{\farcs}3 (Jura \& Werner 1999).  In HD 179821, the weak 12 ${\mu}$m emission results from few grains being near the star.  
In the case of OH 238.1+4.2, there are a large number of grains near the star
yet the material is relatively cold.

\section {THE UNRESOLVED CENTRAL SOURCE}
The infrared emission from OH 231.8+4.2 depends in part upon the composition and size of the grains.  Here, we first consider the possibility that the grains are
sufficiently small 
 that the emissivity of the particles varies
as ${\nu}^{+1}$ and that the envelope is optically thin.  In this case, the average temperature of the grains,
$T_{gr}$ can be estimated implicitly from the formula:
\begin{equation}
\frac{F_{\nu}(11.7 {\mu}m)}{F_{\nu}(17.9 {\mu}m)}\;=\;\left(\frac{{\nu}_{2}}{{\nu}_{1}}\right)^{4}\;\frac{e^{h{\nu}_{1}/kT_{gr}}\,-\,1}{e^{h{\nu}_{2}/kT_{gr}}\,-\,1}
\end{equation}
where ${\nu}_{2}$ and ${\nu}_{1}$ correspond to 11.7 ${\mu}$m and 17.9 ${\mu}$m, respectively.  With $\frac{F_{\nu}(11.7 {\mu}m)}{F_{\nu}(17.9 {\mu}m)}$ = 0.15, then $T_{gr}$ ${\approx}$ 120 K.  If the grains act like black bodies, then $T_{gr}$ ${\approx}$ 135 K is a better fit to the data. 

If the material is opaque, then the observed flux, F$_{\nu}$, is given by the expression:
\begin{equation}
F_{\nu}\;=\;B_{\nu}(T_{gr})\,{\Omega}_{source}
\end{equation}
where $B_{\nu}$ denotes the Planck function and ${\Omega}_{source}$ the solid angle of the source.  
Since the flux at 17.9 ${\mu}$m from the innermost
0{\farcs}5 is 60 Jy, the minimum black body temperature of the grains
which emit this radiation is 130 K. 

In the simplest of optically thin models (see Sopka et al. 1985), if the grain opacity varies as ${\nu}^{+p}$, then the temperature of the dust
grains, $T_{gr}$,  at a distance $D$ from the star can be computed from
the expression:
\begin{equation}
T_{gr}\;=\;T_{*}\,(\frac{R_{*}}{2D})^{\frac{2}{4\,+\,p}}
\end{equation} 
Given that OH 231.8+4.2 has spectral type M9, we adopt $T_{eff}$ = 2500 K
(van Belle et al. 1996).  With an average value of $L$ = 1.5 ${\times}$ 10$^{4}$ L$_{\odot}$, then $R_{*}$ =
4.6 ${\times}$ 10$^{13}$ cm. Since the central source is unresolved, we expect that all the emission from the point source
arises from a region within 0{\farcs}25 of the star which corresponds to  a distance of 4.9 ${\times}$ 10$^{15}$ cm.  At this distance, the
predicted grain temperature from equation (3) for $p$ = 1 (a crude averaging of the silicate opacity given by David \& Pegourie 1995) is 290 K, much greater than the  value inferred from the data.  Even if the particles are black
bodies with $p$ = 0, then the grain temperature is predicted to equal
170 which is larger than the inferred value.

A better way to understand the data is to presume  a nonspherical circumstellar envelope where much of the mass is largely confined to a  disk. Knapp, Sandell \& Robson (1993) showed that
a flat opaque disk cannot explain the spectral energy distribution from OH 231.8+4.2 because in the infrared, such a disk should exhibit F$_{\nu}$ varying as ${\nu}^{0.33}$ while, in fact, between 12 ${\mu}$m and 60 ${\mu}$m, the IRAS data indicate that F$_{\nu}$ varies as ${\nu}^{-2.1}$.
Below, we consider a flared disk which, when viewed face-on, exhibits F$_{\nu}$ varying as ${\nu}^{-1.67}$ (Chiang \& Goldreich 1997).  We suggest that emission from a flared disk at a substantial inclination  can account for the data.    

We can make an estimate of the minimum dust mass in the compact source, $M_{dust}$.  If it were optically thin, then
\begin{equation}
M_{dust}\;=\;\frac{F_{\nu}\,D_{*}^{2}}{B_{\nu}(T_{gr}){\chi}_{\nu}}
\end{equation}
where $D_{*}$ denotes the distance to the source.
If, as is likely, the system is opaque, then this relationship only provides a lower bound to $M_{dust}$. With  $T_{gr}$ = 130 K, F$_{\nu}$(17.9 ${\mu}$m) =
60 Jy and ${\chi}_{\nu}$ = 1000 cm$^{2}$ g$^{-1}$ (Ossenkopf, Henning \& Mathis 1992), then $M_{dust}$ ${\geq}$  7 ${\times}$ 10$^{28}$ g.

\section{A MODEL FOR THE DISK EMISSION}
In order to explain the spectral energy distribution of the unresolved source,
we present a  model to explain the observations of OH 231.8+4.2 which
is based on the calculations by Chiang \& Goldreich
(1997) for  passive, opaque disks around pre-main sequence stars.  Although we consider a post-main sequence system,  the basic physical
principles can be applied to both sorts of disks.   In this  model, the heating and cooling rates
locally balance each other, and the task is to determine the temperature 
which achieves this effect. We ignore the illumination of the disk by light from the bipolar lobes.

If the disk has a local temperature, $T_{disk}$, then in a flat system which is optically thick in the vertical direction, for $D$ $>>$ $R_{*}$,
\begin{equation}
T_{disk}\;{\approx}\;\left(\frac{2}{3{\pi}}\right)^{1/4}\;\left(\frac{R_{*}}{D}\right)^{3/4}T_{*}
\end{equation}
For a disk in vertical hydrostatic equilibrium which is locally isothermal, the density distribution, ${\rho}$, is given by the expression
\begin{equation}
{\rho}\;=\;{\rho}_{0}\,exp\left(-\frac{z^{2}}{h^{2}}\right)
\end{equation}
where
\begin{equation}
h\;=\; \left(\frac{2\,D^{3}\,k_{B}\,T_{disk}}{G\,M_{*}\,{\mu}}\right)^{1/2} 
\end{equation}
and ${\mu}$ is the mean molecular weight.  In this expression, we ignore the self-gravity of the disk.

The transition from flat to  flared disk occurs at a critical distance, $D_{crit}$, which is established by the criterion
that $h$ ${\geq}$ $R_{*}$.  From the above, 
\begin{equation}
D_{crit}\;=\;\left(\frac{3{\pi}}{32}\right)^{1/9}\left(\frac{GM_{*}{\mu}}{k_{B}T_{*}R_{*}}\right)^{4/9}\;R_{*}
\end{equation}
and at this location, the temperature, $T_{crit}$, is given by the expression:
\begin{equation}
T_{crit}\;=\;\left(\frac{256}{81\,{\pi}^{4}}\right)^{1/12}\,\left(\frac{k_{B}T_{*}R_{*}}{GM_{*}{\mu}}\right)^{1/3}\,T_{*}
\end{equation}
Using the stellar  parameters for OH 231.8+4.2 described above, assuming that $M_{*}$ = 1 M$_{\odot}$ because the star has lost mass during its post-main sequence evolution, and assuming that the gas is primarily H$_{2}$ and He so that ${\mu}$ = 3.9 ${\times}$ 10$^{-24}$ g, then $(2k_{B}T_{*}R_{*})/(GM_{*}{\mu})$ = 0.061.  Consequently, $D_{crit}$ = 3.0 $R_{*}$ so that the
flaring of the disk begins close to the star.  At $D$ = $D_{crit}$, the  disk
temperature, $T_{crit}$, is 740 K. 

In their models for  flared disks, Chiang \& Goldreich (1997) define the ``grazing angle", ${\alpha}$, as the angle  between the surface of the disk and the line of sight to the star.  Far from the star, 
\begin{equation}
{\alpha}\;{\approx}\;D\,\frac{d}{dD}\left(\frac{h}{D}\right)
\end{equation}
In this approximation,
\begin{equation}
T_{disk}\;=\;\left(\frac{{\alpha}}{2}\right)^{1/4}\,\left(\frac{R_{*}}{D}\right)^{1/2}\,T_{*}
\end{equation}
With these expressions, the disk temperature is given by the expression:
\begin{equation}
T_{disk}\;=\;\left(\frac{1}{7}\right)^{2/7}\left(\frac{R_{*}}{D}\right)^{3/7}\,\left(\frac{2k_{B}T_{*}R_{*}}{GM_{*}{\mu}}\right)^{1/7}\;T_{*}
\end{equation}

Chiang \& Goldreich (1997) discuss models for pre-main sequence stars whose disks result from the collapse of interstellar clouds.  Such disks might
have very large amounts of angular momentum and extend to great distances from the star.  Here, we picture a disk created in a binary system, and the
initial angular momentum of the system limits the amount of angular
momentum, $J$, that the disk can possess.  Below, we argue that the disk
has an angular momentum given by:
\begin{equation}
J\;=\;M_{disk}\sqrt{\frac{GM_{*}D_{out}}{{\pi}}}
\end{equation}
where $D_{out}$ is the outer boundary of the disk which we set equal
to 3 $v_{vis} t$ where these parameters are defined below (see equation 18). If OH 231.8+4.2 had, itself, an initial mass of 3 M$_{\odot}$ and a companion of 1 M$_{\odot}$ in a circular orbit of radius 
between 3 and 5 AU, then the companion would have had an orbital angular momentum of ${\sim}$2 ${\times}$ 10$^{53}$ g cm$^{2}$ s$^{-1}$.  If half of this orbital angular momentum has been transferred to the disk so that $J$ ${\sim}$ 10$^{53}$ g cm$^{2}$ s$^{-1}$, and if $M_{disk}$ ${\sim}$ 0.1 M$_{\odot}$, we find from equation (13) that $D_{out}$ = 6 ${\times}$ 10$^{15}$ cm, consistent with the observational upper limit that
$D_{out}$ ${\leq}$ 5 ${\times}$ 10$^{15}$ cm.  According to our models,  where $D_{out}$ = 5 ${\times}$ 10$^{15}$ cm, the inferred
disk temperature from equation  (12) is 130 K and $h$ = 3 ${\times}$ 10$^{15}$ cm.

  If the disk were nearly face-on-then we would detect emission from essentially every annular ring.  However,
 if the angle between the normal to the disk
and the line of sight, ${\theta}$, is greater than 45$^{\circ}$, then the inner,
warm portions of the disk are occulted (Chiang \& Goldreich 1999).    According to Kastner et al. (1992) and Shure et al. (1995), ${\theta}$ ${\approx}$ 54$^{\circ}$, and therefore most of the disk is
shadowed from our point of view.  As a first approximation, we assume that we only observe emission from the outer portion of the flared disk which has radius, $D_{out}$,  and a uniform temperature, $T_{out}$.  This outer material  
is so ``flared" into our line of sight that its emission dominates the
unresolved central  source that we detect.  If
${\Omega}_{out}$ denotes the solid angle subtended by the outer portion of
the flared disk from our perspective, then
\begin{equation}
{\Omega}_{out}\;{\approx}\;4\,\left(\frac{D_{out}}{D_{*}}\right)^{2}\,sin\,{\theta}
\end{equation}
We can use equation (2) to estimate the flux from the source.  

The comparison of the model with the data is shown in Figure 4 where we present the results for both $D_{out}$ = 3 ${\times}$ 10$^{15}$ cm (and $T_{out}$ = 130) and an alternative  model with $D_{out}$ = 3 ${\times}$ 10$^{15}$ cm (and $T_{out}$ = 160 K).
The model with $D_{out}$ = 5 ${\times}$ 10$^{15}$ cm  agrees  with our observations of the compact emission at 11.7 ${\mu}$m and 17.9 ${\mu}$m
to better than a  factor of 2.  Our model does not reproduce the total flux from the system since our observations show that the bipolar lobes  contribute the majority of the flux from OH 231.8+4.2 at 17.9 ${\mu}$m and 11.7 ${\mu}$m.  The results from the model  suggest that much 
of the millimeter and submillimeter continuum are produced by the compact source.  This is possible but unproven.  We do not know the millimeter-wavelength opacity of the particles around OH 231.8+4.2, and the disk may not be fully opaque at these wavelengths from the perspective of the Earth.  In fact, the model significantly overestimates the flux at the longest
observed wavelength of about 3 mm. The apparent agreement between the model and many of 
the integrated millimeter and submillimeter fluxes  
shown in Figure 4 may be a coincidence. Also, at 4.9 ${\mu}$m, the observed DIRBE flux is 16 Jy which agrees with
ground-based measurements (Woodward et al. 1989) and, furthermore, this is the total flux expected from the photosphere.  The map of Woodward et al. (1989) shows that
the emission at 4.7 ${\mu}$m is extended and much of it is probably
scattered radiation. The amount of disk thermal emission at wavelengths
shortward of 11.7 ${\mu}$m is not easily measured and is therefore
not included in Figure 4.        
         
\section{DISK DYNAMICS}

The model proposed here is geometrically similar to  that proposed by Cohen et al. (1985). What is uncertain is whether the disk material is expanding radially or whether it is orbiting; the
expansion rate is not well measured.  It has usually been assumed that the
disk is expanding radially at the characteristic outflow speed of 20 km s$^{-1}$.  At this speed, the material reaches the inferred value of $D_{out}$ of 5 ${\times}$ 10$^{15}$ cm in approximately 80 yr.   
If, however, the disk is in vertical hydrostatic equilibrium  then  the vertical sound crossing time must be  short compared to the expansion time.  In the outermost region, the disk temperature may be ${\sim}$100 K and therefore possess a sound speed near 1 km s$^{-1}$.  Since the vertical displacement approaches 5 ${\times}$ 10$^{15}$ cm, the time to achieve vertical hydrostatic equilibrium and thus the implied  disk lifetime is at least ${\sim}$2 ${\times}$ 10$^{3}$ years.   It is easier to understand the presence of  a flared disk if  the material is
orbiting.

Pringle (1991) has presented a model for  how viscous disks around binary stars may expand
as angular momentum is transfered from the binary stars into the disk, and   here we follow his model.  In the standard treatment of accretion disks,
matter flows inwards while angular momentum flows outwards.  Pringle (1991)
argues that for a circumbinary disk, the binary system can provide enough
torque to the disk so that no matter flows inwards, but the outward flow
of angular momentum still occurs.  In this model, after the disk is formed, its mass remains constant.  The torque operates as the binary
creates a tide in the circumbinary disk (Lin \& Papaloizou 1979).        

 We assume that the circumstellar matter is injected into the system as a narrow ring of mass, $M_{disk}$, and an inner radius, $D_{in}$.   The ring then expands as angular momentum is transferred from inner to outer regions by viscosity, ${\nu}$, which can be written as
\begin{equation}
{\nu}\;=\;\frac{2\,{\alpha}\,c_{S}^{2}}{3\,{\Omega}}
\end{equation}
Here, ${\alpha}$ is the usual dimensionless parameter and is taken to be ${\leq}$ 1,
$c_{S}$ is the speed of sound or (${\gamma}k_{B}T_{disk}/{\mu}$)$^{1/2}$ where ${\gamma}$ is the ratio of specific heats, and ${\Omega}$ is the local angular velocity of the disk, $(GM_{*})^{1/2}D^{-3/2}$. 

Following Pringle (1991), we make the approximation that the gas temperature falls as $D^{-1/2}$ so that
\begin{equation}
T_{disk}\;=\;T_{in}\,\left(\frac{D_{in}}{D}\right)^{1/2}
\end{equation}
where $T_{in}$ is the temperature at the inner boundary of the disk, $D_{in}$.  This radial variation of the temperature is somewhat different from that described in equation (12), but it should not lead to any essential change in the physics of the outflow. It does allow for a simple analytic solution to the  asymptotic disk evolution.  Following Pringle, with this radial variation of the temperature,  we define  $v_{vis}$ [denoted $k$ by Pringle] as a characteristic speed of the outflow induced by the viscosity, such that $v_{vis}$ = ${\nu}/D$,  and is given by the expression:
\begin{equation}
v_{vis}\;=\;\frac{2\,{\gamma}\,k_{B}\,T\,{\alpha}}{3{\mu}}\left(\frac{D}{GM_{*}}\right)^{1/2}
\end{equation}
Because $T$ scales as $D^{-1/2}$,  $v_{vis}$ is independent of $D$.
 
The inner radius of the disk might be  ${\sim}$1.7 times greater than the orbital separation of the binary pair (Artymowicz \& Lubow 1994). Although uncertain, we adopt $D_{in}$ = 3 $R_{*}$ = 1.3 ${\times}$ 10$^{14}$ cm where  $T_{disk}$ = 740 K (see above). The value of 
${\alpha}$ is not well known for the material around OH 231.8+4.2.  Saslaw (1978) has shown that grains  around a variable star move in non-circular orbits, and  these non-circular motions might produce   a substantial  viscosity.   
We assume that ${\alpha}$ = 1 implying that $v_{vis}$ = 0.29 km s$^{-1}$. 

After the initial conditions damp out,  Pringle's model for the outer
portion of the disk has a surface density given by the expression:
\begin{equation}
{\Sigma}\;{\approx}\;\frac{M_{disk}}{(12\,{\pi}^{3}\,D^{3}\,v_{vis}\,t)^{1/2}}\;
exp\left(-\frac{D}{3\,v_{vis}\,t}\right)
\end{equation}
where $M_{disk}$ is the mass of the disk.  This expression for the surface density of the disk appears to be independent of the torque being exerted on the disk because the transfer of angular momentum from the binary to the disk is effected by employing the inner boundary condition of the solution for the disk evolution that no matter flows inwards even though matter and angular momentum flow outwards.  That is, at all times,  the mass of the disk is constant and:
\begin{equation}
\int_{0}^{\infty}2\,{\pi}\,D\,{\Sigma}(D)\,dD\;=\;M_{disk}
\end{equation}
  The disk angular momentum, $J$, is given by:
\begin{equation}
J(t)\;=\;M_{disk}\sqrt{\frac{3GM_{*}v_{vis}t}{{\pi}}}
\end{equation}
From equation (18), we write that $D_{out}$ ${\approx}$ 3 $v_{vis}\,t$. 

According to Pringle (1991), the characteristic time for the
torque to operate on the disk, $t_{torque}$, at the innermost radius, is given  by the expression:
\begin{equation}
t_{torque}\;=\;\frac{4D_{in}^{2}}{3{\nu}_{in}}\;=\;\frac{2\,{\mu}{\sqrt{GM_{*}D_{in}}}}{{\alpha}{\gamma}k_{B}T_{in}}
\end{equation} 
Thus, for the disk around OH 231.8+4.2, this characteristic time for the torque to operate is ${\sim}$ 200 years.  The outflow time is $D_{out}/v_{vis}$ or about
5000 yr.  Therefore, the time for the torque to be effected is considerably less than the inferred
outflow time. 
  Alcolea et al. (2001) derive a minimum age of the bipolar outflow
from OH 231.8+4.2 of 770 years. Thus the time noted here for the flared
disk to expand to its current estimated size is comparable to the minimum lifetime of the system.

\section{DISCUSSION}
  We have suggested that the circumstellar disk around OH 231.8+4.2 is
gravitationally bound and orbiting the system rather than simply undergoing
a radial expansion.   Underlying this picture is that OH 231.8+4.2 is an AGB star with a hitherto-unseen companion.
Ignace, Cassinelli \& Bjorkman (1996) have
suggested that mass loss from a red giant with a sufficiently large rate of rotation -- perhaps induced by a companion (Livio \& Soker 1988) -- could produce  a  disk   material within a few stellar radii of the mass-losing star.  Whether this disk is bound or unbound -- that is, whether it is orbiting or expanding  -- is not easily calculated because the acceleration law of the material near the star is not known.   In fact, the acceleration may be time-variable over several pulsational cycles (Winters et al. 2000).  Although it is possible that an orbiting ring could be formed near a rotating
AGB star, this is unproven.
Once such a ring is formed, as discussed above, 
   Pringle (1991) has proposed that the material  could expand
by the viscous dissipation of angular momentum contained within the orbiting
central binary.      

Although there are a large number of studies of the molecular gas around OH 231.8+4.2, there is no direct kinematic demonstration from observations of the circumstellar gas that the disk around OH 231.8+4.2 is orbiting.
Most of the gas emission is dominated by material in the bipolar flow which subtends a solid angle of ${\sim}$ 10$^{-8}$ ster.  The flared disk probably only subtends a solid angle of ${\sim}$ 10$^{-11}$ ster.  Therefore, the disk
emission is mostly overwhelmed by the extended emission from the bipolar flow, and
in order to detect the disk, interferometry is required.

One molecule which has been extensively studied at high angular resolution is OH.
A remarkable aspect of OH 231.8+4.2 is that the centroid of the OH maser 
emission is offset about 1{\arcsec} to the Southwest from the SiO and H$_{2}$O masers
and the millimeter continuum (Gomez \& Rodriguez 2001, Zijlstra et al. 2001,
Sanchez-Contreras et al. 1998). One possibility is that since OH is produced from the photodissociation of H$_{2}$O (Glassgold 1996),   a one-sided ultraviolet flux might produce a one-sided concentration of OH, as might explain  the CO profile toward HD 188037
(Jura et al. 1997). Since M46 with its concentration of ultraviolet-emitting A-type stars also lies to the Southwest of OH 231.8+4.6,   we suggest that the displacement of the OH masers from the core of  OH 231.8+4.6 may be a consequence of its position in the outer, Northeast  portion of M46.      

Alcolea et al. (2001) report interferometric maps of the CO emission with resolution as high as 1{\farcs}5 ${\times}$ 0{\farcs}7.  They
distinguish between the extended bipolar flow and a central core.  Even the
central core displays both the north-south outflow plus a central equatorial disk  or torus.  The innermost disk or torus is not well mapped in their data, and it
is not possible to determine whether it is orbiting or expanding.  

Sanchez Contreras et al. (2000) report interferometric observations of HCO$^{+}$, SO, H$^{13}$CN and SiO around OH 231.8+4.2.  The SO is particularly sensitive
to high density regions, but the angular resolution of the maps 4{\farcs}2 
${\times}$ 2{\farcs}2 is insufficient to distinguish an orbiting disk
of radius 5 ${\times}$ 10$^{15}$ cm. Sanchez Contreras et al. (2000) report an expanding disk or ring of radius 2 ${\times}$ 10$^{16}$ cm at speed of 6-7 km s$^{-1}$.  The relationship between this observed large
disk of 2 ${\times}$ 10$^{16}$ cm radius and the orbiting disk of 5 ${\times}$ 10$^{15}$ cm radius that we propose to exist is unclear.  Interferometric observations of
the circumstellar H$_{2}$O masers report two velocity peaks separated by
0{\farcs}1 in the North-South direction (Gomez \& Rodriguez 2001).  This
orientation suggests that this emission may be excited in the bipolar flow
rather than the disk.  

On a scale of 0{\farcs}005, Desmurs et al. (2001) report
SiO maser emission which is in the East-West direction and with kinematics which are consistent with being
 infalling material at a speed of ${\sim}$10 km s$^{-1}$ with  some additional rotation at a speed, $V_{rot}$, of ${\sim}$6 km s$^{-1}$.  The inferred rotation is measured over a spatial scale of  approximately the diameter of the star.  Unless the system is nearly face-on, the apparent rotation is too small to be  orbital motion; it must reflect  rotation of the star.  The total angular momentum of 
an AGB star is about $\frac{2}{9}M_{*}R_{*}V_{rot}$ (Soker 1998); OH 231.8+4.2 may currently possess ${\sim}$ 10$^{52}$ g cm$^{2}$ s$^{-1}$ of angular momentum.  Above, we suggest that the disk may possess 
${\sim}$ 10$^{53}$ g cm$^{2}$ s$^{-1}$ of angular momentum.  Thus, it is imaginable that there might be enough angular momentum in the system to account for the proposed disk. 

 We conclude that
 the available interferometric data are consistent with the hypothesis of an orbiting disk with an outer radius of 5 ${\times}$ 10$^{15}$ cm, but they do not lend any support to the model.
The argument for the existence of an
orbiting rather than expanding disk is indirect.  While the predicted infrared emission from a  flared, orbiting disk agrees with our data,   we cannot rule out other models for the compact, mid-infrared emission.    

The  mass of the  disk around OH 231.8+4.2 is uncertain. Above we derived a minimum mass of dust of 10$^{29}$ g which would correspond to a total of about 0.01 M$_{\odot}$ if the gas to dust ratio is ${\sim}$160.  The maximum possible dust mass is  about 1.0 M$_{\odot}$ since this would require that most
of the material lost by OH 231.8+4.2 was trapped in the disk.  However, a disk of this mass could not be both orbiting and as large as 5 10$^{15}$ cm because it would possess too much angular momentum.  It is possible that the disk around OH 231.8+4.2 has about 0.1 M$_{\odot}$.  In this case,  the  disk mass around OH 231.8+4.2 is
larger than that for disks around other evolved red giants except, perhaps, the Red
Rectangle (see Jura \& Kahane 1999). With time, the
circumbinary disk around OH 231.8+4.2 must evolve.  If, in fact, the material is orbiting, then
macroscopic solids and even planets might eventually form.
\section{CONCLUSIONS}

We have obtained 11.7 ${\mu}$m and 17.9 ${\mu}$m images of OH 231.8+4.2;  we identify an unresolved central  source with F$_{\nu}$(17.9 ${\mu}$m) = 60 Jy.   We propose that this emission results from  a flared disk with an outer temperature of 130 K and an outer disk radius of 5 ${\times}$ 10$^{15}$ cm.   
 One possible model that is consistent with the data is that this flared disk is orbiting
rather than simply expanding.

This work has been partly supported by  NASA.  We thank J. Alcolea, E. Chiang, M. Morris and N. Soker for very helpful comments.  
   
\newpage
\begin{center}
{\bf FIGURE CAPTIONS}
\end{center}
Fig. 1.  A histogram of $L_{\nu}$(60 ${\mu}$m) for the ``very-dusty" AGB
stars within ${\sim}$1 kpc of the Sun listed by Jura \& Kleinmann (1989).
The distances are taken from that paper and we mostly used  IRAS data for the 60 ${\mu}$m fluxes.  
  For both the Egg Nebula and
OH 231.8+4.2, we take $L_{\nu}$(60 ${\mu}$m) from the time-averaged COBE data.
\\
\\
Fig. 2.  The 11.7 ${\mu}$m image of OH 231.8+4.2.  North is up and East is to the left.  The contour levels (Jy arcsec$^{-2}$) are shown in the color bar.
\\
\\
Fig. 3.  The 17.9 ${\mu}$m image of OH 231.8+4.2.  North is up and East is to the left.  The contour levels (Jy arcsec$^{-2}$) are shown in the color bar.
The very faint emission to the South of the image is an artifact.
\\
\\
Fig. 4.  The spectral energy distribution of OH 231.8+4.2 compared with the model described in the text for an opaque, flared disk.  The color of the squares denotes their sources: blue (IRAS), red (COBE), cyan (IRAM, 
Sanchez Contreras et al. 1998, 2000),   green (JCMT, Knapp et al. 1993) and   black (Kuiper Airborne Observatory, Harvey et al. 1991).  The crosses show our results reported in this paper; black for the fluxes from the unresolved point source and red for the fluxes from the entire source.
The solid black line shows the model described in the text which is comprised of the emission from a black body 
at 130 K from a cylinder of radius 5 ${\times}$ 10$^{15}$ cm at a tilt angle of 54$^{\circ}$.  The dashed line
shows an alternative model with a cylinder at 160 K and a radius of
3 ${\times}$ 10$^{15}$ cm at the same tilt angle of 54$^{\circ}$.     
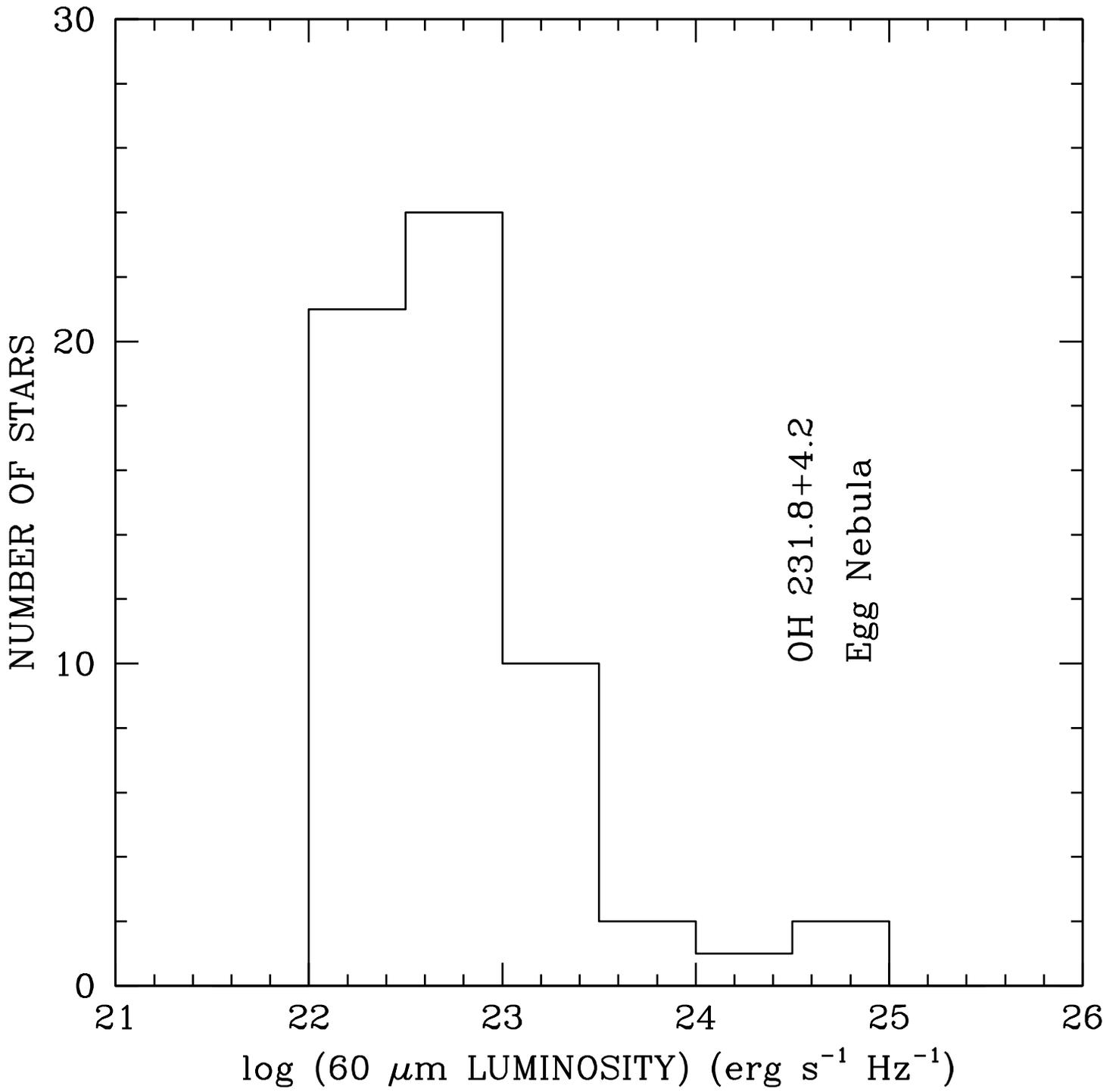
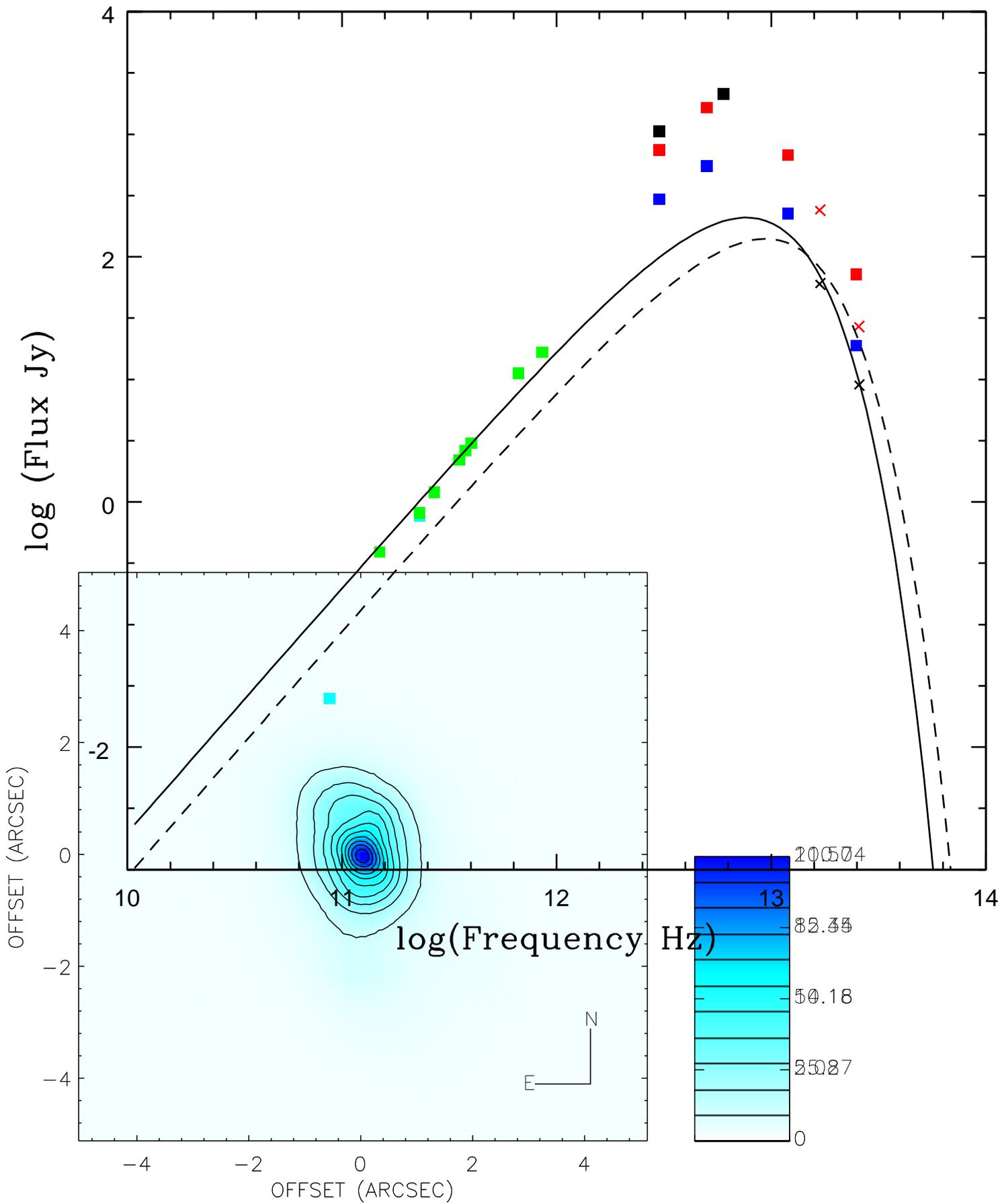
\end{document}